\documentclass[11pt]{article}\textwidth 6.5in\textheight 9in
\usepackage{amssymb}\usepackage[colorlinks]{hyperref}\usepackage{color}
\usepackage{stmaryrd}\usepackage{mathrsfs}
\usepackage{graphicx}
\usepackage{amsmath}
\usepackage{caption}
\usepackage{subcaption}
\usepackage{listings}
\usepackage{verbatim}
\usepackage{braket}

\begin{document}
	\setlength{\captionmargin}{27pt}
	\newcommand\hreff[1]{\href {http://#1} {\small http://#1}}
	\newcommand\trm[1]{{\bf\em #1}} \newcommand\emm[1]{{\ensuremath{#1}}}
	\newcommand\prf{\paragraph{Proof.}}\newcommand\qed{\hfill\emm\blacksquare}
	
	\newtheorem{thr}{Theorem} 
	\newtheorem{lmm}{Lemma}
	\newtheorem{cor}{Corollary}
	\newtheorem{con}{Conjecture} 
	\newtheorem{prp}{Proposition}
	\newcommand\Ip{\I_\mathrm{Prob}}
	\newcommand\QC{\mathbf{QC}} 
	\newcommand\C{\mathbf{C}} 
	\renewcommand\H{\mathbf{H}}
	
	\newtheorem{blk}{Block}
	\newtheorem{dff}{Definition}
	\newtheorem{asm}{Assumption}
	\newtheorem{rmk}{Remark}
	\newtheorem{clm}{Claim}
	\newtheorem{exm}{Example}
	
	\newcommand\Ks{\mathbf{Ks}} 
	\newcommand{\ab}{a\!b}
	\newcommand{\yx}{y\!x}
	\newcommand{\yux}{y\!\underline{x}}
	
	\newcommand\floor[1]{{\lfloor#1\rfloor}}\newcommand\ceil[1]{{\lceil#1\rceil}}

	\newcommand{\bmu}{\boldsymbol{\mu}}
	
	\newcommand{\lea}{<^+}
	\newcommand{\gea}{>^+}
	\newcommand{\eqa}{=^+}

	\newcommand{\lel}{<^{\log}}
	\newcommand{\gel}{>^{\log}}
	\newcommand{\eql}{=^{\log}}
	
	\newcommand{\F}{\mathbf{F}}
	\newcommand{\E}{\mathbf{E}}
	\newcommand{\lem}{\stackrel{\ast}{<}}
	\newcommand{\gem}{\stackrel{\ast}{>}}
	\newcommand{\eqm}{\stackrel{\ast}{=}}
	
	\newcommand\edf{{\,\stackrel{\mbox{\tiny def}}=\,}}
	\newcommand\edl{{\,\stackrel{\mbox{\tiny def}}\leq\,}}
	\newcommand\then{\Rightarrow}

	\newcommand\ml{\underline{\mathbf m}}
	
	\renewcommand\chi{\mathcal{H}}
	\newcommand\km{{\mathbf {km}}}\renewcommand\t{{\mathbf {t}}}
	\newcommand\KM{{\mathbf {KM}}}\newcommand\m{{\mathbf {m}}}
	\newcommand\md{{\mathbf {m}_{\mathbf{d}}}}\newcommand\mT{{\mathbf {m}_{\mathbf{T}}}}
	\newcommand\K{{\mathbf K}} \newcommand\I{{\mathbf I}}
	
	\newcommand\II{\hat{\mathbf I}}
	\newcommand\Kd{{\mathbf{Kd}}} \newcommand\KT{{\mathbf{KT}}} 
	\renewcommand\d{{\mathbf d}} 
	\newcommand\D{{\mathbf D}}
	\newcommand\Tr{\mathrm{Tr}}
	\newcommand\w{{\mathbf w}}
	\newcommand\Cs{\mathbf{Cs}} \newcommand\q{{\mathbf q}}
	\newcommand\St{{\mathbf S}}
	\newcommand\M{{\mathbf M}}\newcommand\Q{{\mathbf Q}}
	\newcommand\ch{{\mathcal H}} \renewcommand\l{\tau}
	\newcommand\tb{{\mathbf t}} \renewcommand\L{{\mathbf L}}
	\newcommand\bb{{\mathbf {bb}}}\newcommand\Km{{\mathbf {Km}}}
	\renewcommand\q{{\mathbf q}}\newcommand\J{{\mathbf J}}
	\newcommand\z{\mathbf{z}}
	
	\newcommand\B{\mathbf{bb}}\newcommand\f{\mathbf{f}}
	\newcommand\hd{\mathbf{0'}} \newcommand\T{{\mathbf T}}
	\newcommand\R{\mathbb{R}}\renewcommand\Q{\mathbb{Q}}
	\newcommand\N{\mathbb{N}}\newcommand\BT{\{0,1\}}
	\newcommand\FS{\BT^*}\newcommand\IS{\BT^\infty}
	\newcommand\FIS{\BT^{*\infty}}
	\renewcommand\S{\mathcal{C}}\newcommand\ST{\mathcal{S}}
	\newcommand\UM{\nu_0}\newcommand\EN{\mathcal{W}}
	
	\newcommand{\supp}{\mathrm{Supp}}
	
	\newcommand\lenum{\lbrack\!\lbrack}
	\newcommand\renum{\rbrack\!\rbrack}
	
	\newcommand\h{\mathbf{h}}
	\renewcommand\qed{\hfill\emm\square}
	\renewcommand\i{\mathbf{i}}
	\newcommand\p{\mathbf{p}}
	\renewcommand\q{\mathbf{q}}
	\title{Two Quantum Paradigms, but Still No Signal}
	
	\author {Samuel Epstein\\samepst@jptheorygroup.org}
	
	\maketitle
	
	\begin{abstract}
		An overwhelming majority of quantum (pure and mixed) states, when undertaking a POVM measurement, will result in a classical probability with no algorithmic information. Thus most quantum states produce white noise when measured.  Furthermore most non-pointer states, when undergoing the decoherence process, will produce white noise. These results can be seen as consequences of the vastness of Hilbert spaces.
	\end{abstract}
	\section{Introduction}
	This paper details the barrier between the classical and quantum realms. We investigate two paradigms. In the Copenhagen approach, quantum physics is relegated to the microscopic world, surrending completeness. Thus the universe is not considered to be a quantum state. When a quantum state interacts with macroscopic measurement apparatus, the wave function collapses, producing a signal. The mathematical notion of the measurement apparatus is a POVM measurement. Another approach is to model the measurement apparatus and the environment as quantum states that can become entangled with the quantum state they are measuring. In this setup, quantum decoherence and einselection becomes essential to describing the measurement process. With einselection, the system-environment interaction preserves an set of pointer sets, which form a preferred orthonormal pointer basis. Thus decoherence and einselection are compatible with PVMs, a subset of POVMs.
	
	This paper shows that in both paradigms, given a measurement apparatus, an overwhelming majority of quantum states will produce \textit{algorithmic garbage}. Thus for most quantum states, the resulting probability produced with respect to a POVM measurement will have no signal. Similarly in decoherence, the non-pointer states will overwhelmingly produce garbage. 
	\section{POVM Measurements}
	Information non-growth laws say information about a target source cannot be increased with randomized processing. In classical information theory, we have \cite{CoverJo91}
	
	$$I(g(X)\,{:}\,Y)\leq I(X\,{:}\,Y),$$
	where $g$ is a randomized function, $X$ and $Y$ are random variables, and $I$ is the mutual information function. Thus processing a channel at its output will not increase its capacity. Information conservation carries over into the algorithmic domain, with the inequalities \cite{Levin84,EpsteinDerandom22}
	$$\I(f(x):y) \lea \I(x:y);\hspace{1.5cm}\I(f(a);\ch) \lea \I(a;\ch).$$
	
	The information function is $\I(x:y)=\K(x)+\K(y)-\K(x,y)$, where $\K$ is Kolmogorov complexity. The other term is $\I(a;\ch)=\K(a)-\K(a|\ch)$, where $\ch\in\IS$ is the halting sequence. These inequalities ensure target information cannot be obtained by processing. If for example the second inequality  was not true, then one can potentially obtain information about $\ch$ with simple functions. Obtaining information about $\ch$ violates the Independence Postulate,  (see \cite{Levin13}). Information non growth laws can be extended to signals \cite{EpsteinProb23} which can be modeled as probabilities over $\N$ or Euclidean space\footnote{In \cite{EpsteinProb23} probabilities over $\IS$ and $T_0$ second countable topologies were also studied.}. The ``signal strength'' of a probability $p$ over $\N$ is measured by its self information.
	$$
	\Ip(p:p) = \log \sum_{i,j}2^{\I(i:j)}p(i)p(j).
	$$
	
	A signal, when undergoing randomized processing $f$,  will lose its cohesion\footnote{A probability $p$, when processed by a channel $f:\FS\times\FS\rightarrow \R_{\geq 0}$ is a new probability $fp(x)=\sum_z f(x|z)p(z)$.}. Thus any signal going through a classical channel will become less coherent \cite{EpsteinProb23}.
	$$
	\Ip(f(p):f(p)) \lea \Ip(p:p).
	$$
	
	In Euclidean space, probabilities that undergo convolutions with probability kernels will lose self information. For example a signal spike at a random position will spread out when convoluted with the Gaussian function, and lose self information. The above inequalities deal with classical transformations. One can ask, is whether, quantum information processing can add new surprises to how information signals occur and evolve. 
	
	One can start with the prepare-and-measure channel, also known as a Holevo-form channel. Alice starts with a random variable $X$ that can take values $\{1,\dots,n\}$ with corresponding probabilities $\{p_1,\dots,p_n\}$. Alice prepares a quantum state, corresponding to density matrix $\rho_X$, chosen from $\{\rho_1,\dots,\rho_n\}$ according to $X$. Bob performs a measurement on the state $\rho_X$, getting a classical outcome, denoted by $Y$. Though it uses quantum mechanics, this is a classical channel $X\rightarrow Y$. So using the above inequality, cohesion will deteriorate regardless of $X's$ probability, with
	Th$$\Ip(Y:Y)\lea\Ip(X:X).$$
	
	There remains a second option, constructing a signal directly from a mixed state. This involves constructing a mixed state, i.e. density matrix $\sigma$, and then performing a POVM measurement\footnote{A POVM measurement $E$ is a collection of positive-semi definite Hermitian matrices $\{E_k\}$ such that $\sum_k E_k=1$. Given a state $\sigma$, $E$ induces a probability over the measurements of the form $E\sigma(k)=\Tr E_k\sigma$.} $E$ on the state, inducing the probability $E\sigma(\cdot)$. However from \cite{EpsteinProb23}, for elementary (even enumerable) probabilities $E\sigma$,
	$$\Ip(E\sigma\,{:}\,E\sigma)\lea \K(\sigma,E).$$
	Thus for simply defined density matrices and measurements, no signal can appear. So experiments that are simple will result in simple measurements, or white noise. However it could be that a larger number of uncomputable pure or mixed states produce coherent signals. However, theorems in \cite{EpsteinAPhysics24} say otherwise, in that given a POVM measurement $E$, a vast majority of pure and mixed states will have negligible self-information. Thus for uniform distributions  $\Lambda$ and $\mu$  over pure and mixed states\footnote{The distribution $\eta$ over mixed states is $\mu(\sum_{i=1}^M p_i\ket{\psi_i}\bra{\psi_i}) = \eta(p_1,\dots,p_M)\prod_{i=1}^M\Lambda(\ket{\psi_i})$, where $\eta$ is any probability over the $M$-simplex.}\footnote{The proof to these inequalities is in the running for the strangest in AIT, relying on a lower computable combination of \textit{upper} computable tests.},
	$$\int 2^{\Ip(E\ket{\psi}:E\ket{\psi})}d\Lambda = O(1);\hspace{1cm}\int 2^{\Ip(E\sigma:E\sigma)}d\mu(\sigma) = O(1).$$
	
	This can be seen as a consequence of the vastness of Hilbert spaces as opposed to the limited discriminatory power of quantum measurements. In addition, there could be non-uniform distributions of pure or mixed states that could be of research interest. 
	
	
	However the measurement process has a surprising consquence, whereas for most states, an initial measurement produces no signal, the subsequent wave function collapse causes a massive uptake in algorithmic signal strength of the states. Thus a second measurement will produce a valid signal. Let $F$ be a PVM\footnote{A PVM measurement is a POVM measurement where the measurement operators are projectors.}, of $2^{n-c}$ projectors, of an $n$ qubit space and let $\Lambda_F$ be the distribution of pure states when $F$ is measured over the uniform distribution $\Lambda$. Thus $\Lambda_F$ represents the $F$-collapsed states from $\Lambda$. Note that if $F$ has two few projectors, it lacks discretionary power to produce a meaningful signal when the states are in distribution $\Lambda_F$. A theorem from \cite{EpsteinAPhysics24} states the following.
	
	$$n-2c\lel\log \int 2^{\Ip(F\ket{\psi}:F\ket{\psi})}d\Lambda_F.$$
	
	\section{Decoherence}
	
	The following letter of Einstein to Born (April 1954) illustrated the problem of superposition of quantum macrosystems.
	\begin{quote}
		\textit{
			Let $\Psi_1$ and $\Psi_2$ be two solutions to the Same Schr\"{o}dinger equation\dots When the system is a macrosystem and when $\Psi_1$ and $\Psi_2$ are `narrow
			with respect to position, then in by far the greater number of cases this is no longer true $\Psi_{12}=\Psi_1+\Psi_2$. Narrowness with respect to macrocoordinates is not only independent of the principles of quantum mechanics, but is, moreover, incompatible with them.}
	\end{quote}
	This letter brings up the astonishing fact that observables on the microscale and absent from everyday experiments. In fact, \textit{quantum decoherence} and \textit{einselection} show that such superpositions are highly fragile and decay exponentially fast. The root cause of this phenomena is caused by interactions between a system and environment. A closed system assumption is a fundamental obstacle to the study of the transition of the quantum domain to the classical domain.
	
	In this light, the setup is a (microscopic) system and (macroscopic) environment. Given joint Hamiltonian dynamics between the system and environment, there are two main consequences.
	\begin{enumerate}
		\item The effective disappearance of coherence, the source of quantum interference effects, from the system.
		\item The dynamical definition of preferred ``pointer states'', which are unchanged by the system/environment dynamics.
	\end{enumerate}
	The phenomena of (1) is called \textit{decoherence} (see \cite{Schlosshauer10} for an extensive overview). The phenomena (2) is called \textit{einselection}, short for Environment INduced Selection \cite{Zurek03}. In Einselection, the system-environment Hamiltonian ``selects'' a set of prefered quasi-classical ``pointer states'' which do not decohere. Einselection explains why we only observe a few ``classical'' quantities such as momentum and positon, and not superpositions of these pointer states.
	
	We begin our explanation with a two state case, which can be generalized to arbitrary number of pointer states. Suppose the system is described by a superposition of two quantum states $\ket{\psi_1}$ and $\ket{\psi_2}$ which for example can be thought of as two localization of two positions $x_1$ and $x_2$ in a double slit experiment. The system/environment interaction results in
	\begin{align*}
		\ket{\psi_1}\ket{E_0}&\rightarrow \ket{\psi_1}\ket{E_1}\\
		\ket{\psi_2}\ket{E_0}&\rightarrow \ket{\psi_2}\ket{E_2}.
	\end{align*}
	So the state of the environment evolves according to the state of the system. Now if the system is in a superposition of $\ket{\psi_1}$ and $\ket{\psi_2}$, we get the dynamics
	$$
	\frac{1}{\sqrt{2}}(\ket{\psi_1}+\ket{\psi_2})\ket{E_0}\rightarrow \frac{1}{\sqrt{2}}(\ket{\psi_1}\ket{E_1}+\ket{\psi_2}\ket{E_2})
	$$
	The reduced density matrix of system (with the environment traced out) is
	$$
	\frac{1}{2}\left(\ket{\psi_1}\bra{\psi_1}+\ket{\psi_2}\bra{\psi_2}+\ket{\psi_1}\bra{\psi_2}\braket{E_2|E_1}+\ket{\psi_2}\bra{\psi_2}\braket{E_1|E_2}\right).
	$$
	The last two terms correspond to the interference between the state $\ket{\psi_1}$ and $\ket{\psi_2}$. If the environment recorded the position of the particle, then $\ket{E_1}$ and  $\ket{E_2}$ will be approximately orthogonal. In fact, it can be shown that in many dynamics, $\braket{E_1|E_2}\leq e^{-t/\tau}$, where $t$ is the time of the interaction and $\tau$ is a positive constant. In this case
	$$
	\rho \approx \frac{1}{2}\left(\ket{\psi_1}\bra{\psi_1}+\ket{\psi_2}\bra{\psi_2}\right).
	$$
	Thus virtually all coherence between the two states $\ket{\psi_1}$ and $\ket{\psi_2}$ is lost. The states $\ket{\psi_1}$ and $\ket{\psi_2}$ are called invariant to the dynamics, and will not undergo decoherence. They are called ``pointer states'' because they induce an apparatus with a pointer mechanism to be orientated at a particular angle. Einselection preserves ``pointer states'' but superpositions of them are fragile and do not survive the dynamics with the system.
	\section{Predictability Sieve}
	In general, there is not a clear division between pointer and non-pointer states. Instead one can use a score to measure how much of the state has been preserved. The interaction of pointer states with the environment is predictable; they are effectively classical states. However a state that is heavily decohered is unpredictable. Let $\ket{\psi}$ be an initial pure state, and $\rho_{\ket{\psi}}(t)$ be the density matrix of the system state after interacting with the environment for time $t$. The loss of predictability caused by the environment can be measured in the following two measures
	\begin{itemize}
		\item $\varsigma^T_{\ket{\psi}}(t) = \Tr \rho_{\ket{\psi}}^2(t).$ 
		\item $\varsigma^S_{\ket{\psi}}(t) = S(\rho_{\ket{\psi}}(t)).$
	\end{itemize}
	The first measure, uses squared trace of the density matrix whereas the second measure uses von Neumann entropy. The first measure will start at 1 and then decrease proportionately to much much the state decoheres. This is similarly true for the von Neumann entropy predictability sieve, except the measure starts at 0. 
	
	In this section we introduce an algorithmic predictability sieve $\varsigma^A$. Assume a basis of $2^n$ pointer states. Let the system be $\ket{\psi}$, an arbitrary pure state. We consider the limit of interacting with the environment as time approaches infinity. In this idealized case, the decoherence $\ket{\psi}\bra{\psi}$ into a classical probability, with the off-diagonal terms turned to 0. Let $p_{\ket{\psi}}$ be the classical probability that $\ket{\psi}$ decoheres to, with $p_{\ket{\psi}}(i)=\ket{\psi}\bra{\psi}_{ii}$. 
	
	\begin{dff}[Algorithmic Predictability Sieve]
		$\varsigma^A(\ket{\psi}) = \Ip(p_{\ket{\psi}}:p_{\ket{\psi}}|n)$.
	\end{dff}
	Thus, $\varsigma^A$ is the self information of the probability measure induced by the diagonal of the density matrix $\ket{\psi}\bra{\psi}$. Note that this self information is relativized to $n$, that is the universal Turing machine $U$ has $n$ on an auxiliary tape. On average, pointer states $\ket{i}$ have high algorithmic predictability. 
	$$ \frac{1}{2^n}\sum_{i=1}^{2^n}\varsigma^A(\ket{i})\eqa n.$$ 
	We now show that an overwhelming majority of pure states over the pointer basis decohere into algorithmic white noise. Due to algorithmic conservation inequalities, there is no (even probabilisitic) method of processing this white noise to produce a signal. Thus superpositions of pointer bases will produce garbage that can't be measured. The following statement shows that almost all pure states decohere into algebraic garbage. 	Let $\Lambda$ be the uniform distribution on the unit sphere of an $n$ qubit space. 
	$$\int 2^{\varsigma^A(\ket{\psi})}d\Lambda = O(1).$$
	Apriori distributions which are close to $\Lambda$ also have this property. Let $\Gamma$ be a distribution over $n$ qubit pure states such that $\Gamma(\ket{\psi})\leq 2^c\Lambda(\ket{\psi})$ for all $\ket{\psi}$.
	$$
	\log \int 2^{\varsigma^A(\ket{\psi})}d\Gamma \lea c.
	$$


\end{document}